\RequirePackage[hyphens]{url}
\documentclass[journal=jpclcd,manuscript=letter]{achemso}

\usepackage[version=3]{mhchem} 
\usepackage{xcolor}
\usepackage{soul} 
\usepackage{amssymb}
\usepackage{lineno,hyperref}
\usepackage{graphicx}
\usepackage{dcolumn}
\usepackage{bm}
\usepackage{float}

\newcommand{\DIPC}[0]{
Donostia International Physics Center (DIPC),
Paseo Manuel de Lardizabal 4, 20018 Donostia-San Sebasti\'an, Spain}
\newcommand{\CFM}[0]{
Centro de F\'{\i}sica de Materiales CFM/MPC (CSIC-UPV/EHU), Paseo Manuel de Lardizabal 5, 20018 Donostia-San Sebasti\'an, Spain}
\newcommand{\POT}[0]{
Institut f\"{u}r Chemie, Universit\"at Potsdam, Karl-Liebknecht-Straße 24-25, D-14476 Potsdam, Germany}

\newcommand{\HITS}[0]{
Heidelberg Institute for Theoretical Studies (HITS gGmbH), Schloss-Wolfsbrunnenweg 35, 69118 Heidelberg (Germany),
and
Interdisciplinary Center for Scientific Computing (IWR), Ruprecht-Karls-Universität Heidelberg, Im Neuenheimer Feld 205, 69120 Heidelberg (Germany)
}

\author{Auguste TETENOIRE}
\email{auguste.tetenoire@dipc.org}
 \affiliation{\DIPC}

\author{Christopher Ehlert}
\email{christopher.ehlert@h-its.org}
 \affiliation{\HITS}

\author{J.\ I.\ Juaristi}
\email{josebainaki.juaristi@ehu.eus}
\affiliation{Departamento de Pol\'{\i}meros y Materiales Avanzados: F\'{\i}sica, Qu\'{\i}mica y Tecnolog\'{\i}a, Facultad de Qu\'{\i}micas (UPV/EHU), Apartado 1072, 20080 Donostia-San Sebasti\'an, Spain}
\alsoaffiliation{\CFM}
\alsoaffiliation{\DIPC}

\author{Peter Saalfrank}
\email{peter.saalfrank@uni-potsdam.de}
 \affiliation{\POT}

\author{M. Alducin}
\email{maite.alducin@ehu.eus}
 \affiliation{\CFM}
 \alsoaffiliation{\DIPC}

\title{Why Ultrafast Photo-induced CO Desorption Dominates over Oxidation on Ru(0001)}

\date{\today}

\begin{document}

\begin{tocentry}

\includegraphics{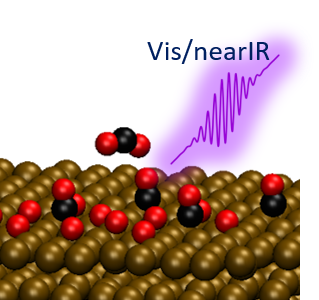}

\end{tocentry}

\begin{abstract}
CO oxidation on Ru(0001) is a long-standing example of a reaction that, being thermally forbidden in ultra-high vacuum, can be activated by femtosecond laser pulses. In spite of its relevance, the precise dynamics of the photo-induced oxidation process as well as the reasons behind the dominant role of the competing CO photo-desorption remain unclear. Here we use ab initio molecular dynamics with electronic friction that account for the highly excited and non-equilibrated system created by the laser to investigate both reactions. Our simulations successfully reproduce the main experimental findings: the existence of photo-induced oxidation and desorption, the large desorption to oxidation branching ratio, and the changes in the O K-edge X-ray absorption spectra attributed to the initial stage of the oxidation process. Now, we are able to monitor in detail the ultrafast CO desorption and CO oxidation occurring in the highly-excited system and to disentangle what causes the unexpected inertness to the otherwise energetically favored oxidation.
\end{abstract}



The behavior of Ru(0001) towards CO oxidation is quite unique. By varying the experimental conditions from ambient to low CO pressures, the Ru(0001) surface changes from the most active to the most inert transition metal surface. In particular, early experimental studies carried out for different mixed coverages of O and CO demonstrated that under ultra high vacuum (UHV) conditions the oxidation process leading to CO$_2$ formation cannot be thermally activated~\cite{Kostov1992Nov}. Interestingly, the reaction can be initiated by irradiating the system with near-infrared femtosecond laser pulses~\cite{bonn99, obergJCP2015}, while thermally only CO desorption is found~\cite{bonn99}. Photons in this energy range are efficiently absorbed by the metal electrons that can subsequently transfer energy to the adsorbates directly and also indirectly, via the excited surface phonons that result from electron-phonon coupling. The dependence of the reaction probabilities on the delay between two correlated pulses suggests that the oxidation events are initiated thanks to direct coupling of the adsorbed O atoms to the laser-excited electrons, thus explaining that the very same reaction cannot be thermally activated. Still, the competing and indirect laser-induced CO desorption mechanism largely dominates over CO$_2$ desorption, with an observed branching ratio of around 35 between CO desorption and CO oxidation. 

As remarkable as it is, the precise dynamics of the oxidation process in this emblematic  experiment remains unknown. In fact, already the reasons behind the very low probability for CO$_2$ formation are unclear because density functional theory (DFT) calculations of the minimum energy path (MEP) of these reactions show that CO$_2$ desorption is energetically favored against CO desorption~\cite{obergJCP2015,tetenoire2021}.  Altogether, the general questions that naturally arise are: How do the ultrafast CO oxidation and CO desorption proceed in the highly-excited and non-equilibrated system? And ultimately, what mechanisms make the photoinduced CO desorption more probable than oxidation?

Here we use ab initio molecular dynamics simulations with electronic friction (AIMDEF) that account for the non-equilibrated excited electrons and phonons created in the irradiated (O,CO)/Ru(0001) system to answer these questions. Our simulations allow us to resolve the precise dynamics of the laser-induced desorption and oxidation processes and understand the experimental findings. The large desorption to oxidation branching ratio, which is successfully reproduced here, is caused by the extremely reduced configurational space leading to oxidation as compared to CO desorption. Furthermore, we additionally use our dynamics simulations to calculate the evolution of the O K-edge X-ray absorption spectra (XAS) during the oxidation process. Our real-time spectra reproduce the spectral changes that were observed in ultrafast pump-probe X-ray spectroscopy experiments~\cite{ostrom2015Feb}, confirming them as fingerprints of the initial stage of the oxidation process. The good agreement achieved here further corroborates the validity of our model to capture and describe the photo-induced oxidation dynamics on the highly excited (O,CO)/Ru(0001) surface.

Photo-induced desorption and oxidation of CO from the (O,CO)-covered Ru(0001) surface is simulated with ab initio classical molecular dynamics using the ($T_\textrm{e}, T_\textrm{l}$)-AIMDEF methodology~\cite{alducin2019} (see also the Supporting Information (SI)~\cite{si}). In particular, the laser-excited electrons and concomitant electron-excited phonons are described within the two temperature model (2TM)~\cite{anisimov74} as two coupled heat thermal baths, which are characterized by time-dependent electron and lattice temperatures $T_\textrm{e}(t)$ and $T_\textrm{l}(t)$, respectively. Next, the effect of the excited electrons in the adsorbates dynamics is included by means of Langevin equations of motion, in which electronic friction and random forces model the coupling of the adsorbates to the electronic thermal bath defined by $T_\textrm{e}(t)$. Furthermore, effects due to the excited phonons are incorporated by coupling the surface atoms in the upper layers to the Nos\'{e}-Hoover thermostat~\cite{Nose84,Hoover1985}, assuring that the lattice temperature evolves as $T_\textrm{l}(t)$. The ($T_\textrm{e}, T_\textrm{l}$)-AIMDEF method, as well as variants of it, have been widely used to simulate the femtosecond laser induced dynamics and reactions of adsorbates at metal surfaces~\cite{tullyss94,springer96,vazha09,fuchsel10,fuchsel11,loncaricprb16, loncaricnimb16, scholz16, juaristiprb17,alducin2019, scholz19}.  The typical short lifetime of the electronic excited states at metal surfaces (of the order of few fs) justifies employing a Langevin description because the dynamics evolves in the ground potential energy surface most of the time~\cite{vazha09,fuchsel10,fuchsel11,saalfrankcr06}. 

The experimental (2O+CO)/Ru(0001) honeycomb surface~\cite{ostrom2015Feb}, in which CO adsorbs atop a Ru atom and the O atoms occupy the second nearest hcp and fcc sites forming a honeycomb arrangement around the CO, is modeled using a periodic slab with five Ru layers and the adlayer (see Figure~S1 in SI). The employed (4$\times$2) surface cell containing two equivalent adsorbates of each kind is the minimum cell that permits including out of phase movements of the adsorbates and a reliable description of the interadsorbates interactions, that are expected to be relevant at sufficiently large coverages~\cite{denzler03,xin15,sung16,juaristiprb17,alducin2019, serrano2021,muzas22}. All ($T_\textrm{e}, T_\textrm{l}$)-AIMDEF simulations are performed with {\sc vasp}~\cite{vasp1,vasp2} and the AIMDEF module~\cite{blanco14,saalfrank14,novkoprb15,novkoprb16,novkonimb16,novkoprb17,juaristiprb17} using the same computational parameters and the same van der Waals exchange-correlation functional by Dion \textit{et al.}~\cite{Dion2004} that were used in our previous structural study~\cite{tetenoire2021}. Electronic friction coefficients are calculated with the local density friction approximation (LDFA)~\cite{juaristi08,alducinpss17}.

The O K-edge XAS have been obtained with the transition potential and $\Delta$-Kohn-Sham methods as implemented in GPAW \cite{enkovaara2010}. Transition probabilities from the O1s core orbitals are calculated from a Fermi’s Golden Rule approach using the core and unoccupied orbitals (a Gaussian broadening of 0.5 eV full-width at half-maximum is used). A shifting method beyond the transition potential method is applied to gauge the lowest-energy peak. For the overall spectrum at a single time step, the approach is applied to all individual oxygen atoms and the final spectrum is obtained by averaging. (For more computational details see SI~\cite{si}.) 
\begin{figure}
\includegraphics*[width=1\columnwidth]{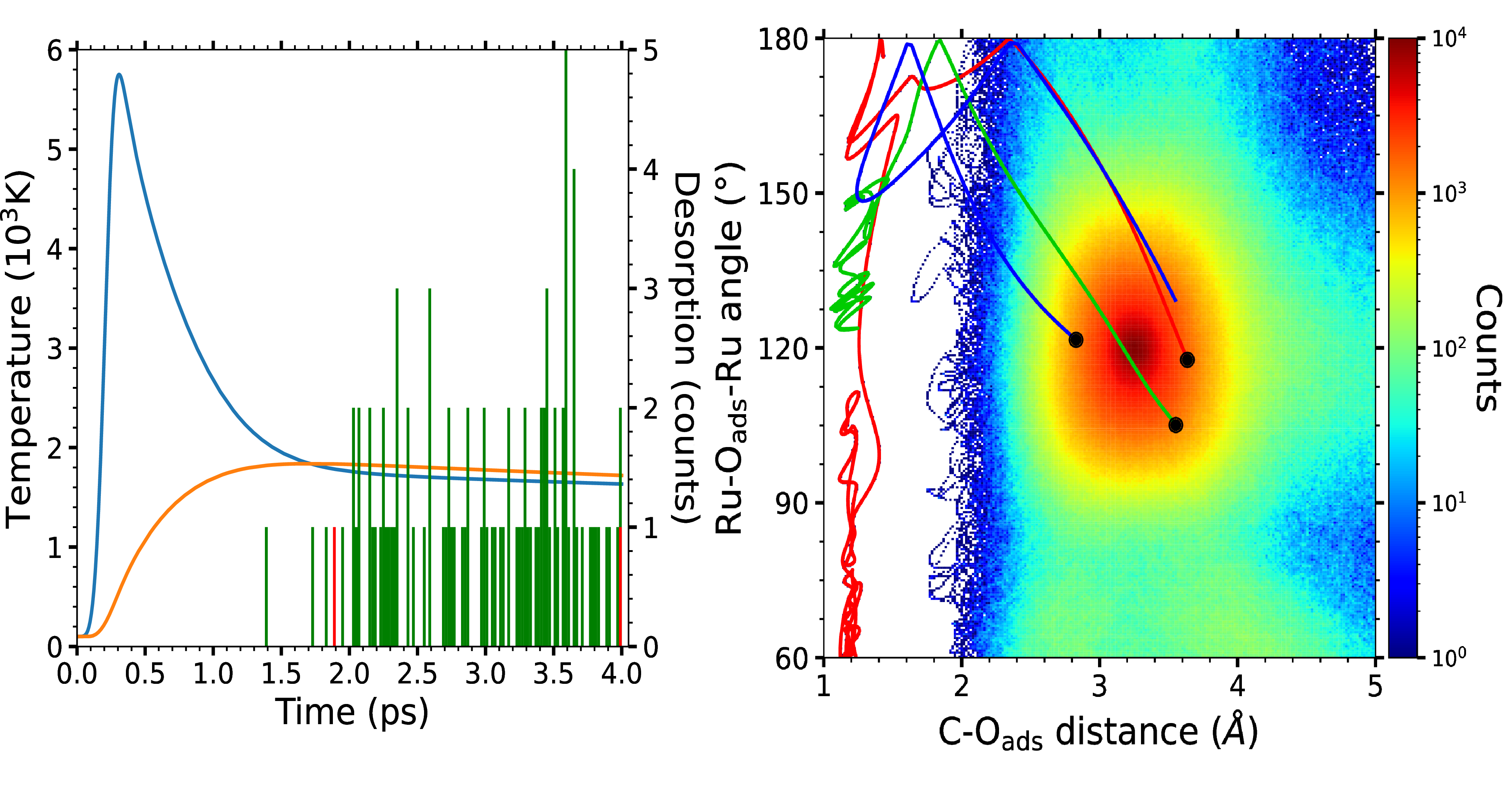}%
\caption{Left: Electronic (blue curve) and phononic (orange curve) temperatures (left $y$-axis) used in the simulations. The peak of the pump laser pulse is at 236~fs. The instants at which the desorption and oxidation events occur are respectively plotted by green and red histograms (right $y$-axis), using 20~fs as bin width. Right: Density plot of the instantaneous in-plane dihedral angle Ru-O$_\textrm{ads}$-Ru and the distance between that O$_\textrm{ads}$ and the C atom in its nearest CO (see SI~\cite{si} for calculation details). For clarity, the values of the two oxidation events are shown by a red and a green line, respectively. The blue line corresponds to the third O adsorbate that reaches the transition state, but it cannot recombine. Only the interval of 50~fs before and after reaching the transition state for oxidation is shown by each line. The black circles correspond to the starting points of these three lines.
\label{fig:2tm}}
\end{figure}

\begin{table*} 
\caption{\label{tab:table1} AIMDEF desorption and oxidation probabilities, P$_\textrm{des}$(CO) and P$_\textrm{des}$(CO$_2$), for (2O+CO)/Ru(0001) and F=200~J/m$^2$. The CO to CO$_2$ branching ratio is compared to the available experimental values~\cite{bonn99,obergJCP2015}.}
\begin{tabular}{cccc}
\hline 
Simulation & P$_\textrm{des}$ (CO) & P$_\textrm{des}$ (CO$_2$) & ratio\\ \hline
($T_{e},T_\textrm{l}$)-AIMDEF & 18.25\% &  0.5\% & 36.5\\
refs~\citenum{bonn99},~\citenum{obergJCP2015} & -- & -- & 35, 31 \\
\hline

\end{tabular}

\end{table*}

Table~\ref{tab:table1} summarizes the results of our simulations for the experimental conditions corresponding to exciting the (2O+CO)/Ru(0001) surface with a 800~nm Gaussian pulse of 110~fs duration and an absorbed fluence F=200~J/m$^2$~\cite{bonn99}. The $T_\textrm{e}(t)$ and $T_\textrm{l}(t)$ curves calculated with the 2TM for these experimental conditions are shown in Figure~\ref{fig:2tm} (input parameters as in refs~\citenum{vazha09,scholz16,juaristiprb17}). We run 200 ($T_\textrm{e}, T_\textrm{l}$)-AIMDEF trajectories with total simulation time of 4~ps each~\cite{parameters}, being the system initially thermalized at 100~K. The desorption (oxidation) probability per molecule is obtained by dividing the total number of CO (CO$_\textrm{2}$) desorbing molecules~\cite{desorption} by the total number of trajectories and the total number of CO in the cell. The result of our ($T_\textrm{e}, T_\textrm{l}$)-AIMDEF simulations is clear. As observed in experiments, after photoexcitation both CO desorption and CO$_2$ formation take place and the former largely dominates over the latter. The corresponding calculated probabilities of 18.25\% and 0.5\% yield a branching ratio between the two processes in nearly perfect agreement with the experimental values P$_\textrm{des}$(CO)/P$_\textrm{des}$(CO$_2$)$=35$~\cite{bonn99} and 31~\cite{obergJCP2015}. The agreement must be considered as qualitative because of the limited statistics we have for the oxidation process (two oxidation events in 200 trajectories) and our integration time of 4~ps (note that some oxidation and desorption events may occur beyond this interval, as found for CO desorption from Pd(111)~\cite{muzas22}), but it remarks already the validity of our non-equilibrium two temperature picture. As shown below, the fact that the XAS features attributed to the oxidation reaction path are also well reproduced makes our simulations all the more reliable. 

\begin{figure*}
\includegraphics*[width=1.0\textwidth]{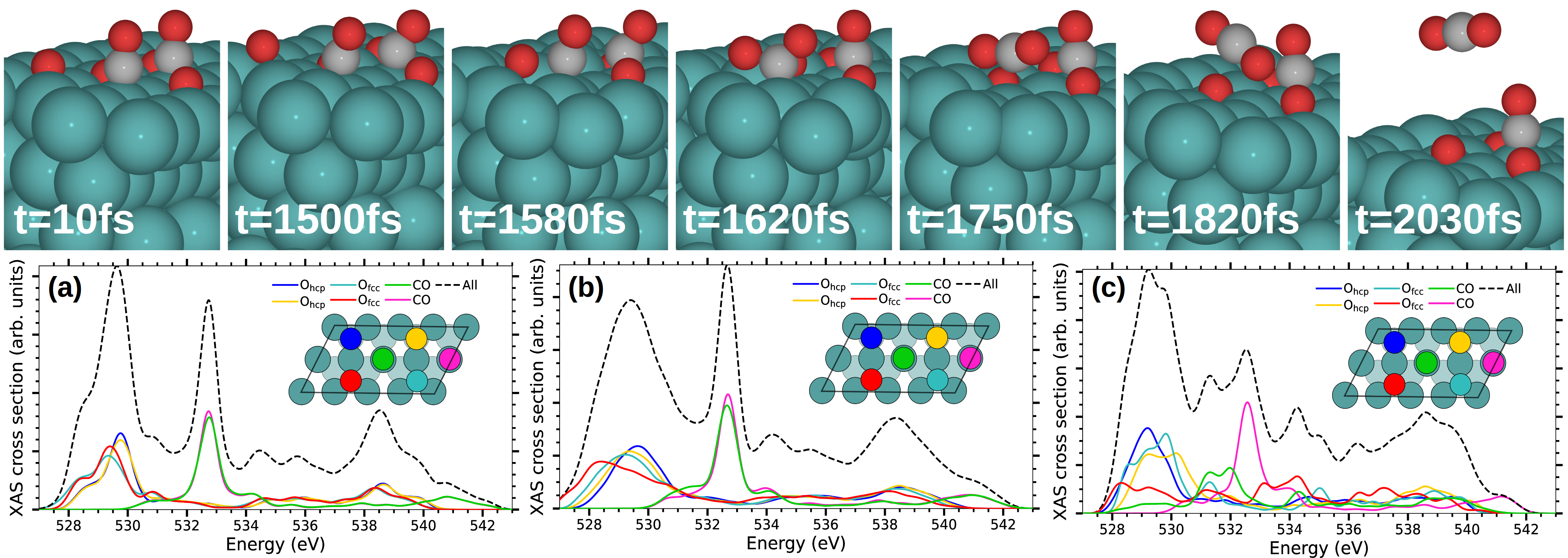}%
\caption{Top: Snapshots of a representative CO oxidation dynamics obtained in the ($T_\textrm{e}, T_\textrm{l}$)-AIMDEF simulations (blue, red, and gray spheres correspond to Ru, O, and C atoms, respectively). The AIMDEF simulation time is indicated in each panel ($t$=0 as in Figure~\ref{fig:2tm}). Bottom: Time averaged O-K XAS cross section calculated at characteristic time intervals during the oxidative desorption process in the selected trajectory: (a) initial strong excitation of the adsorbates (0-1250~fs), (b) access to the transition state ruling the reaction, in which the recombining O$_\textrm{fcc}$ reaches the bridge site that separates it from the nearest CO (1250-1580~fs), and (c) formation of the chemisorbed bent CO$_2$ (1600-1630~fs). Each color curve shows the contribution of the corresponding colored O atom depicted in the surface unit cell plotted as an inset. Black dashed curves show the total time-averaged XAS.
\label{fig:snapshots}}
\end{figure*}

The results of the ($T_\textrm{e}, T_\textrm{l}$)-AIMDEF simulations provide us with a detailed understanding of the laser-induced oxidation process. The fact that CO desorption dominates over CO oxidation is particularly striking considering that previous DFT calculations of the minimum energy reaction paths  showed that desorption requires around 0.38~eV of additional energy compared to oxidation at this coverage~\cite{tetenoire2021}. The natural question that arises is whether the laser-induced oxidation dynamics does or does not proceed through that minimum energy oxidation path, in which the less bound O$_\textrm{fcc}$ abandons its adsorption well, crosses the bridge site between two Ru atoms, and recombines with the nearby CO that tilts to form the chemisorbed bent CO$_2$ (hereafter denoted bCO$_2$). The selected snapshots depicted in Figure~\ref{fig:snapshots} for one of the trajectories confirm that this is the case. In both trajectories leading to CO$_2$ desorption, it is an O$_\textrm{fcc}$ adsorbate that recombines with CO. During the first picosecond upon arrival of the laser pulse, all the adsorbates become highly vibrationally excited. Thus, the CO molecules, although bound to the Ru atom below, tilt profoundly, while the O$_\textrm{fcc}$ and O$_\textrm{hcp}$ adsorbates explore the upper part of the wells, approaching the bridge site regions. It is in the interval $t\simeq$1580-1600~fs that the recombining O$_\textrm{fcc}$ crosses the bridge site and the chemisorbed bCO$_2$ is formed at $t\simeq$1600-1630~fs. From the chemisorbed state the molecule evolves towards the physisorbed linear CO$_2$ ($t\simeq$1820~fs) and finally desorbs at $t\gtrsim$2~ps. Altogether, the figure remarks the complexity of the oxidation process as compared to the simpler CO desorption dynamics, which do not involve intermediate states and barriers~\cite{tetenoire2021}. By analyzing the details of the dynamics, we conclude that it is the reduced configurational space of the former, and particularly the access to the transition state, that explains why the energetically less favorable CO desorption process dominates. This is inferred from the density plot of Figure~\ref{fig:2tm} showing the distribution of the instantaneous Ru-O$_\textrm{fcc,hcp}$-Ru (in-plane) dihedral angles and the distance between that adsorbed O and the C atoms in the two nearest CO (see SI~\cite{si}). A successful recombination requires that the adsorbed O reaches the bridge site (characterized by a dihedral angle of 180$^{\circ}$) and encounters a CO slightly displaced towards the ahead hollow site and adequately tilted to facilitate the C-O bonds (C-O$_\textrm{fcc,hcp}$ distances smaller than 2~{\AA}). These are the properties defining the transition state. As shown in the figure, the access to the bridge site is rather probable, but only on three of these occurrences the  excited O finds a CO correctly oriented for recombination. The fact that only two of the three trajectories shown in the figure end as desorbing CO$_2$ shows that the existence of recrossing events in the transition state makes the oxidation process even more difficult. It is also remarkable to observe that the full oxidation process, which unavoidably involves various intermediate states, lasts a few picoseconds. This is explicitly shown in Figure~\ref{fig:2tm}, where the instants for desorption and oxidation are plotted together with the time evolution of the electronic and lattice temperatures~\cite{time}. In summary, not only the energetics determines reactivity, but the dynamics plays an important role, too.

Our dynamics simulations confirm the interpretation provided in ref~\citenum{ostrom2015Feb} on the changes observed in the time-resolved XAS experiments. However, that interpretation was based on static DFT calculations under the assumption, questionable under non equilibrium conditions, that the oxidation process follows the minimum energy reaction path with the surface at equilibrium. This approach fully neglects that the reaction proceeds in an extremely dynamically perturbed system. Our dynamics simulations are free of those assumptions and treat explicitly the highly excited environment. As a consequence, they provide an explanation of the measured time-dependent XAS and permit to unravel the mechanisms that govern them. The calculated O K-edge XAS is shown in Figs.~\ref{fig:snapshots}(a)-(c) at selected time intervals characterizing the initial stages in the oxidation dynamics followed by the trajectory shown in the same figure. In agreement with experiments, the absorption peak at $\sim$530~eV associated to the recombining O$_\textrm{fcc}$ shifts and broadens towards lower energies as the adsorbate approaches the Ru--Ru bridge site [compare panels (a) and (b)]. Few tens of fs later, the absorption peak assigned to the CO $2\pi^*$ excitation ($\sim$533~eV) red shifts as the distance between the recombining  O$_\textrm{fcc}$ and CO decreases and start to form the chemisorbed bCO$_2$ [compare panel (c) to (a) and (b)]. Also in agreement to experiments, at this oxidation stage we observe that the initial wide weak peak at $\sim$539~eV, which is equally contributed by the four O adsorbates [panel (a)], is transformed into a broad structure that extends from about 536 to 540~eV. Comparison of panels (a) and (b) to (c) shows that the change is caused by the reacting O$_\textrm{fcc}$ (compare the red curves).
%
\begin{figure}[h]
\includegraphics[width=1.0\columnwidth]{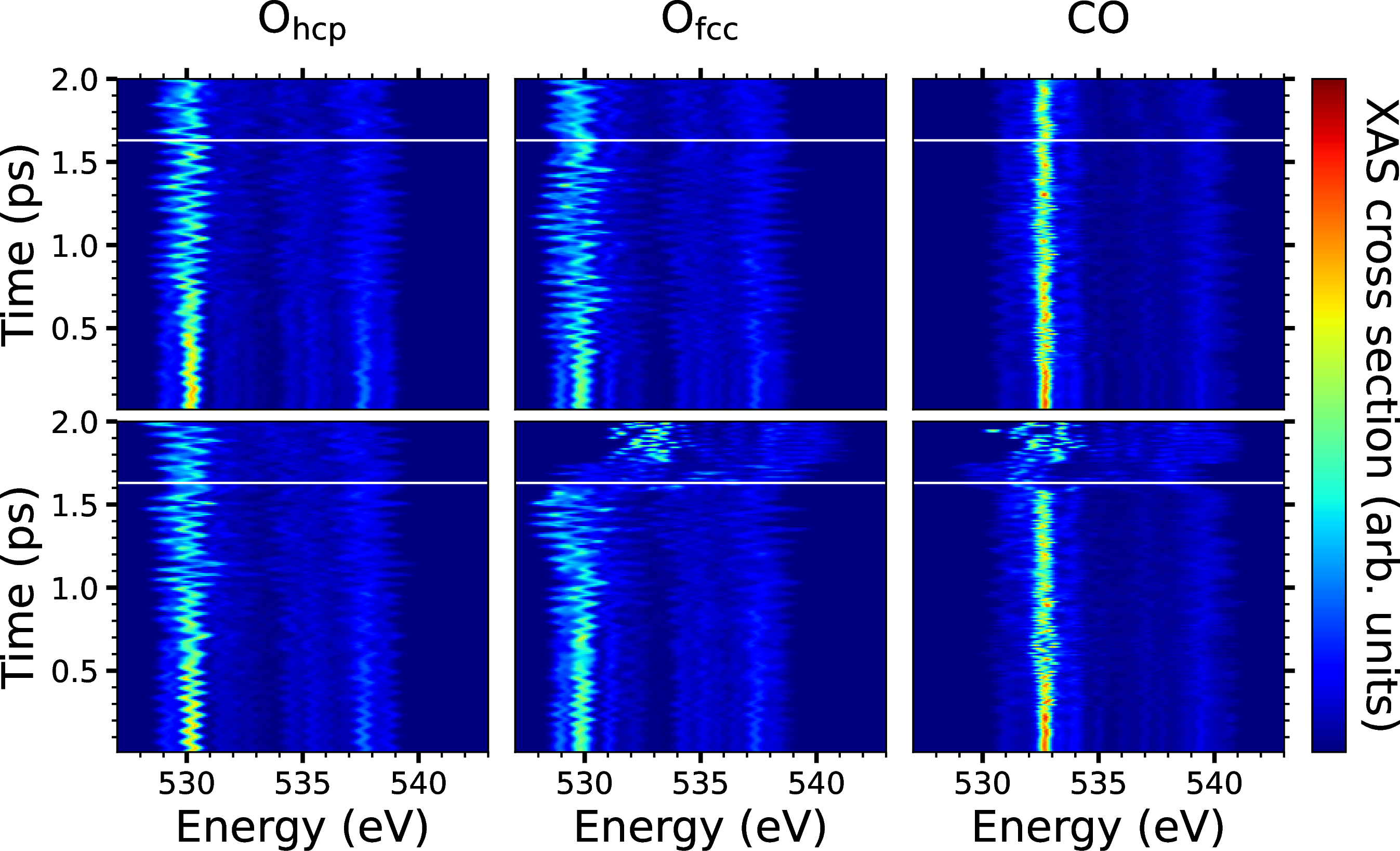}%
\caption{Instantaneous O-K XAS of the trajectory shown in Figure~\ref{fig:snapshots}, resolved for each O atom in the simulation cell: O$_\textrm{hcp}$ (left panels), O$_\textrm{fcc}$ (middle panels), and CO (right panels), with the recombining O$_\textrm{fcc}$ and CO being at the bottom. For illustrative purposes, the first instant at which the O$_\textrm{fcc}$-C distance matches the bCO$_2$ internuclear distance (1.14~{\AA}) is shown by a white line in each plot.
\label{fig:xas2d}}
\end{figure}

The time-resolved O K-edge XAS spectra plotted in Figure~\ref{fig:xas2d} for each adsorbate in the simulation cell allow us to track in detail the contribution of each adsorbate to the changes observed in the spectra associated to the trajectory of Figure~\ref{fig:snapshots}. During the first initial femtosecond, the dominant absorption peaks of O$_\mathrm{hcp}$ and O$_\mathrm{fcc}$ are slightly shifted from each other because of their minor different chemical environment. In all cases, the main absorption peak evolves during the first hundreds of fs following a zigzag structure that reflects the small displacements experienced by the adsorbates as a consequence of the electronic and phononic excitations.  The observed red (blue) shifts in both O$_\mathrm{hcp}$ and O$_\mathrm{fcc}$ correlate with the instants at which adsorbates approach (separate from) the nearby Ru atoms. During the interval lasting from about 1.25 to 1.6~ps, the recombining O$_\mathrm{fcc}$ approaches and successfully reaches the bridge position that separates it from the adsorbed CO. This stage is identified in its corresponding spectra by a profound net red shift of the dominant absorption peak from $\sim$530~eV to $\sim$528~eV, that is not observed in any of the other O adsorbates. The comparatively small changes (oscillations) in the two CO spectra underlines that the molecules remain firmly bound atop the Ru atom, except for the minor displacements associated to frustrated rotations and translations. At 1.6~ps the absorption peak of the recombining CO vanishes, while a new broaden and less intense peak appears in the range 531-532~eV. This new structure is ascribed to the starting point in the formation of the chemisorbed bCO$_2$. The additional analysis of the partial density of states and partial electron densities allows us to confirm the formation of all the bCO$_2$ orbitals at 1.63~ps (see Figure~S4 in SI). The nascent molecule stays about 100~fs in this state before definitely breaking the C-Ru bond and reach the physisorbed linear CO$_2$. Although experimentally inaccessible, note that the absorption spectra in the physisorption state are basically identical for the O atoms, as expected. 

In conclusion, our ab initio molecular dynamics simulations that describe the laser-excited system in terms of non-equilibrated time-dependent electronic and phononic temperatures have successfully reproduced the main features observed in femtosecond laser experiments performed on Ru(0001) covered with a mixed adlayer of O and CO: (i) the photo-desorption of both CO and CO$_2$, (ii) the large branching ratio between desorption and oxidation that exceeds one order of magnitude and (iii) the changes in the O K-edge X-ray absorption spectra that were associated to the initial stage of the oxidation. Thanks to these simulations we can monitor in detail the elementary steps of the desorption and the oxidation dynamics promoted by the laser and determine the reaction paths in the excited system that explain why CO desorption dominates over the energetically favored oxidation. It is the O adsorbed at fcc sites that primarily recombines with the adsorbed CO, following basically the intermediate extreme states of the minimum energy oxidation path. The reason behind the unexpected inertness to the otherwise energetically favored oxidation is twofold: (i) the difficult access to the transition state region, that requires the O atom crossing the bridge site and finding the CO conveniently close and tilted to form the chemisorbed bent CO$_2$ and (ii) the fact that this access does not guarantee a successful recombination. 

\begin{acknowledgement}
A.T, J.I.J, and M.A. acknowledge financial support by the Gobierno Vasco-UPV/EHU [Project No. IT1569-22] and by the Spanish MCIN/AEI/10.13039/501100011033 [Grant No. PID2019-107396GB-I00]. P.S. acknowledges support by the Deutsche Forschungsgemeinschaft (DFG), through project Sa 547-18. C.E. acknowledges the Klaus Tschira Foundation for financial support. This research was conducted in the scope of the Transnational Common Laboratory (LTC) “QuantumChemPhys – Theoretical Chemistry and Physics at the Quantum Scale”. Computational resources were provided by the DIPC computing center.
\end{acknowledgement}

\begin{suppinfo}
The Supporting Information is available free of charge at XXX and includes further information on the ($T_e$,$T_l$)-AIMDEF and XAS simulations.
\end{suppinfo}


\bibliography{refs-acs}

\providecommand{\noopsort}[1]{}\providecommand{\singleletter}[1]{#1}%
\providecommand{\latin}[1]{#1}
\providecommand*\mcitethebibliography{\thebibliography}
\csname @ifundefined\endcsname{endmcitethebibliography}
  {\let\endmcitethebibliography\endthebibliography}{}
\begin{mcitethebibliography}{42}
\providecommand*\natexlab[1]{#1}
\providecommand*\mciteSetBstSublistMode[1]{}
\providecommand*\mciteSetBstMaxWidthForm[2]{}
\providecommand*\mciteBstWouldAddEndPuncttrue
  {\def\EndOfBibitem{\unskip.}}
\providecommand*\mciteBstWouldAddEndPunctfalse
  {\let\EndOfBibitem\relax}
\providecommand*\mciteSetBstMidEndSepPunct[3]{}
\providecommand*\mciteSetBstSublistLabelBeginEnd[3]{}
\providecommand*\EndOfBibitem{}
\mciteSetBstSublistMode{f}
\mciteSetBstMaxWidthForm{subitem}{(\alph{mcitesubitemcount})}
\mciteSetBstSublistLabelBeginEnd
  {\mcitemaxwidthsubitemform\space}
  {\relax}
  {\relax}

\bibitem[Kostov \latin{et~al.}(1992)Kostov, Rauscher, and
  Menzel]{Kostov1992Nov}
Kostov,~K.~L.; Rauscher,~H.; Menzel,~D. {Adsorption of CO on Oxygen-covered
  Ru(001)}. \emph{Surf. Sci.} \textbf{1992}, \emph{278}, 62--86\relax
\mciteBstWouldAddEndPuncttrue
\mciteSetBstMidEndSepPunct{\mcitedefaultmidpunct}
{\mcitedefaultendpunct}{\mcitedefaultseppunct}\relax
\EndOfBibitem
\bibitem[Bonn \latin{et~al.}(1999)Bonn, Funk, Hess, Denzler, Stampfl,
  Scheffler, Wolf, and Ertl]{bonn99}
Bonn,~M.; Funk,~S.; Hess,~C.; Denzler,~D.~N.; Stampfl,~C.; Scheffler,~M.;
  Wolf,~M.; Ertl,~G. Phonon- Versus Electron-Mediated Desorption and Oxidation
  of {CO} on {Ru(0001)}. \emph{Science} \textbf{1999}, \emph{285},
  1042--1045\relax
\mciteBstWouldAddEndPuncttrue
\mciteSetBstMidEndSepPunct{\mcitedefaultmidpunct}
{\mcitedefaultendpunct}{\mcitedefaultseppunct}\relax
\EndOfBibitem
\bibitem[\"Oberg \latin{et~al.}(2015)\"Oberg, Gladh, Marks, Ogasawara, Nilsson,
  Pettersson, and \"Ostr\"om]{obergJCP2015}
\"Oberg,~H.; Gladh,~J.; Marks,~K.; Ogasawara,~H.; Nilsson,~A.; Pettersson,~L.
  G.~M.; \"Ostr\"om,~H. Indication of Non-thermal Contribution to Visible
  Femtosecond Laser-Induced {CO} Oxidation on {Ru(0001)}. \emph{J. Chem. Phys}
  \textbf{2015}, \emph{143}, 074701\relax
\mciteBstWouldAddEndPuncttrue
\mciteSetBstMidEndSepPunct{\mcitedefaultmidpunct}
{\mcitedefaultendpunct}{\mcitedefaultseppunct}\relax
\EndOfBibitem
\bibitem[Tetenoire \latin{et~al.}(2021)Tetenoire, Juaristi, and
  Alducin]{tetenoire2021}
Tetenoire,~A.; Juaristi,~J.~I.; Alducin,~M. Insights into the Coadsorption and
  Reactivity of {O and CO} on {Ru(0001)} and Their Coverage Dependence.
  \emph{J. Phys. Chem. C} \textbf{2021}, \emph{125}, 12614--12627\relax
\mciteBstWouldAddEndPuncttrue
\mciteSetBstMidEndSepPunct{\mcitedefaultmidpunct}
{\mcitedefaultendpunct}{\mcitedefaultseppunct}\relax
\EndOfBibitem
\bibitem[{\ifmmode\ddot{O}\else\"{O}\fi}str{\ifmmode\ddot{o}\else\"{o}\fi}m
  \latin{et~al.}(2015){\ifmmode\ddot{O}\else\"{O}\fi}str{\ifmmode\ddot{o}\else\"{o}\fi}m,
  {\ifmmode\ddot{O}\else\"{O}\fi}berg, Xin, LaRue, Beye, Dell'Angela, Gladh,
  Ng, Sellberg, Kaya, Mercurio, Nordlund, Hantschmann, Hieke,
  K{\ifmmode\ddot{u}\else\"{u}\fi}hn, Schlotter, Dakovski, Turner, Minitti,
  Mitra, Moeller, F{\ifmmode\ddot{o}\else\"{o}\fi}hlisch, Wolf, Wurth, Persson,
  N{\o}rskov, Abild-Pedersen, Ogasawara, Pettersson, and
  Nilsson]{ostrom2015Feb}
{\ifmmode\ddot{O}\else\"{O}\fi}str{\ifmmode\ddot{o}\else\"{o}\fi}m,~H.;
  {\ifmmode\ddot{O}\else\"{O}\fi}berg,~H.; Xin,~H.; LaRue,~J.~b.; Beye,~M.~b.;
  Dell'Angela,~M.; Gladh,~J.; Ng,~M.~L.; Sellberg,~J. A.~a.; Kaya,~S.
  \latin{et~al.}  {Probing the Transition State Region in Catalytic {CO}
  Oxidation on {Ru}}. \emph{Science} \textbf{2015}, \emph{347}, 978--982\relax
\mciteBstWouldAddEndPuncttrue
\mciteSetBstMidEndSepPunct{\mcitedefaultmidpunct}
{\mcitedefaultendpunct}{\mcitedefaultseppunct}\relax
\EndOfBibitem
\bibitem[Alducin \latin{et~al.}(2019)Alducin, Camillone, Hong, and
  Juaristi]{alducin2019}
Alducin,~M.; Camillone,~N.; Hong,~S.-Y.; Juaristi,~J.~I. Electrons and Phonons
  Cooperate in the Laser-Induced Desorption of {CO} from {Pd(111)}. \emph{Phys.
  Rev. Lett.} \textbf{2019}, \emph{123}, 246802\relax
\mciteBstWouldAddEndPuncttrue
\mciteSetBstMidEndSepPunct{\mcitedefaultmidpunct}
{\mcitedefaultendpunct}{\mcitedefaultseppunct}\relax
\EndOfBibitem
\bibitem[si()]{si}
See the Supporting Information for more details on the ($T_e$,$T_l$)-AIMDEF and
  XAS simulations.\relax
\mciteBstWouldAddEndPunctfalse
\mciteSetBstMidEndSepPunct{\mcitedefaultmidpunct}
{}{\mcitedefaultseppunct}\relax
\EndOfBibitem
\bibitem[Anisimov \latin{et~al.}(1974)Anisimov, Kapeliovich, and
  Perel'man]{anisimov74}
Anisimov,~S.~I.; Kapeliovich,~B.~L.; Perel'man,~T.~L. Electron Emission from
  Metal Surfaces Exposed to Ultrashort Laser Pulses. \emph{Sov. Phys.-JETP}
  \textbf{1974}, \emph{39}, 375\relax
\mciteBstWouldAddEndPuncttrue
\mciteSetBstMidEndSepPunct{\mcitedefaultmidpunct}
{\mcitedefaultendpunct}{\mcitedefaultseppunct}\relax
\EndOfBibitem
\bibitem[Nos\'e(1984)]{Nose84}
Nos\'e,~S. A Unified Formulation of the Constant Temperature Molecular Dynamics
  Methods. \emph{J. Chem. Phys.} \textbf{1984}, \emph{81}, 511--519\relax
\mciteBstWouldAddEndPuncttrue
\mciteSetBstMidEndSepPunct{\mcitedefaultmidpunct}
{\mcitedefaultendpunct}{\mcitedefaultseppunct}\relax
\EndOfBibitem
\bibitem[Hoover(1985)]{Hoover1985}
Hoover,~W.~G. Canonical Dynamics: Equilibrium Phase-space Distributions.
  \emph{Phys. Rev. A} \textbf{1985}, \emph{31}, 1695--1697\relax
\mciteBstWouldAddEndPuncttrue
\mciteSetBstMidEndSepPunct{\mcitedefaultmidpunct}
{\mcitedefaultendpunct}{\mcitedefaultseppunct}\relax
\EndOfBibitem
\bibitem[Springer \latin{et~al.}(1994)Springer, Head-Gordon, and
  Tully]{tullyss94}
Springer,~C.; Head-Gordon,~M.; Tully,~J.~C. Simulations of Femtosecond
  Laser-Induced Desorption of {CO} from {Cu(100)}. \emph{Surf. Sci.}
  \textbf{1994}, \emph{320}, L57--L62\relax
\mciteBstWouldAddEndPuncttrue
\mciteSetBstMidEndSepPunct{\mcitedefaultmidpunct}
{\mcitedefaultendpunct}{\mcitedefaultseppunct}\relax
\EndOfBibitem
\bibitem[Springer and Head-Gordon(1996)Springer, and Head-Gordon]{springer96}
Springer,~C.; Head-Gordon,~M. Simulations of The Femtosecond Laser-Induced
  Desorption of {CO} from {Cu(100)} At 0.5 {Ml} Coverage. \emph{Chem. Phys.}
  \textbf{1996}, \emph{205}, 73 -- 89\relax
\mciteBstWouldAddEndPuncttrue
\mciteSetBstMidEndSepPunct{\mcitedefaultmidpunct}
{\mcitedefaultendpunct}{\mcitedefaultseppunct}\relax
\EndOfBibitem
\bibitem[Vazhappilly \latin{et~al.}(2009)Vazhappilly, Klamroth, Saalfrank, and
  Hernandez]{vazha09}
Vazhappilly,~T.; Klamroth,~T.; Saalfrank,~P.; Hernandez,~R. Femtosecond-Laser
  Desorption of {${\mathrm{H}}_{2} ({\mathrm{D}}_{2})$} from {Ru(0001)}:
  Quantum and Classical Approaches. \emph{J. Phys. Chem. C} \textbf{2009},
  \emph{113}, 7790--7801\relax
\mciteBstWouldAddEndPuncttrue
\mciteSetBstMidEndSepPunct{\mcitedefaultmidpunct}
{\mcitedefaultendpunct}{\mcitedefaultseppunct}\relax
\EndOfBibitem
\bibitem[F\"{u}chsel \latin{et~al.}(2010)F\"{u}chsel, Klamroth, Tremblay, and
  Saalfrank]{fuchsel10}
F\"{u}chsel,~G.; Klamroth,~T.; Tremblay,~J.~C.; Saalfrank,~P. Stochastic
  Approach to Laser-Induced Ultrafast Dynamics: The Desorption of
  {${\mathrm{H}}_{2}/{\mathrm{D}}_{2}$} from {Ru(0001)}. \emph{Phys. Chem.
  Chem. Phys.} \textbf{2010}, \emph{12}, 14082--14094\relax
\mciteBstWouldAddEndPuncttrue
\mciteSetBstMidEndSepPunct{\mcitedefaultmidpunct}
{\mcitedefaultendpunct}{\mcitedefaultseppunct}\relax
\EndOfBibitem
\bibitem[F\"{u}chsel \latin{et~al.}(2011)F\"{u}chsel, Klamroth, Monturet, and
  Saalfrank]{fuchsel11}
F\"{u}chsel,~G.; Klamroth,~T.; Monturet,~S.; Saalfrank,~P. Dissipative Dynamics
  within the Electronic Friction Approach: The Femtosecond Laser Desorption of
  {${\mathrm{H}}_{2}/{\mathrm{D}}_{2}$} from {Ru(0001)}. \emph{Phys. Chem.
  Chem. Phys.} \textbf{2011}, \emph{13}, 8659--8670\relax
\mciteBstWouldAddEndPuncttrue
\mciteSetBstMidEndSepPunct{\mcitedefaultmidpunct}
{\mcitedefaultendpunct}{\mcitedefaultseppunct}\relax
\EndOfBibitem
\bibitem[Lon\ifmmode \check{c}\else \v{c}\fi{}ari\ifmmode~\acute{c}\else
  \'{c}\fi{} \latin{et~al.}(2016)Lon\ifmmode \check{c}\else
  \v{c}\fi{}ari\ifmmode~\acute{c}\else \'{c}\fi{}, Alducin, Saalfrank, and
  Juaristi]{loncaricprb16}
Lon\ifmmode \check{c}\else \v{c}\fi{}ari\ifmmode~\acute{c}\else \'{c}\fi{},~I.;
  Alducin,~M.; Saalfrank,~P.; Juaristi,~J.~I. Femtosecond-Laser-Driven
  Molecular Dynamics on Surfaces: Photodesorption of Molecular {Oxygen} from
  {Ag(110)}. \emph{Phys. Rev. B.} \textbf{2016}, \emph{93}, 014301\relax
\mciteBstWouldAddEndPuncttrue
\mciteSetBstMidEndSepPunct{\mcitedefaultmidpunct}
{\mcitedefaultendpunct}{\mcitedefaultseppunct}\relax
\EndOfBibitem
\bibitem[Lon\v{c}ari\'{c} \latin{et~al.}(2016)Lon\v{c}ari\'{c}, Alducin,
  Saalfrank, and Juaristi]{loncaricnimb16}
Lon\v{c}ari\'{c},~I.; Alducin,~M.; Saalfrank,~P.; Juaristi,~J.~I. Femtosecond
  Laser Pulse Induced Desorption: a Molecular Dynamics Simulation. \emph{Nucl.
  Instrum. Methods B} \textbf{2016}, \emph{382}, 114 -- 118, The 21st
  International Workshop on Inelastic Ion Surface Collisions (IISC-21)\relax
\mciteBstWouldAddEndPuncttrue
\mciteSetBstMidEndSepPunct{\mcitedefaultmidpunct}
{\mcitedefaultendpunct}{\mcitedefaultseppunct}\relax
\EndOfBibitem
\bibitem[Scholz \latin{et~al.}(2016)Scholz, Flo\ss{}, Saalfrank, F\"uchsel,
  Lon\ifmmode \check{c}\else \v{c}\fi{}ari\ifmmode~\acute{c}\else \'{c}\fi{},
  and Juaristi]{scholz16}
Scholz,~R.; Flo\ss{},~G.; Saalfrank,~P.; F\"uchsel,~G.; Lon\ifmmode
  \check{c}\else \v{c}\fi{}ari\ifmmode~\acute{c}\else \'{c}\fi{},~I.;
  Juaristi,~J.~I. Femtosecond-Laser Induced Dynamics of {CO} On {Ru(0001)}:
  Deep Insights from a Hot-Electron Friction Model Including Surface Motion.
  \emph{Phys. Rev. B.} \textbf{2016}, \emph{94}, 165447\relax
\mciteBstWouldAddEndPuncttrue
\mciteSetBstMidEndSepPunct{\mcitedefaultmidpunct}
{\mcitedefaultendpunct}{\mcitedefaultseppunct}\relax
\EndOfBibitem
\bibitem[Juaristi \latin{et~al.}(2017)Juaristi, Alducin, and
  Saalfrank]{juaristiprb17}
Juaristi,~J.~I.; Alducin,~M.; Saalfrank,~P. Femtosecond Laser Induced
  Desorption of {${\mathrm{H}}_{2},{\mathrm{D}}_{2}$}, and {HD} from
  {Ru(0001)}: Dynamical Promotion and Suppression Studied with Ab Initio
  Molecular Dynamics with Electronic Friction. \emph{Phys. Rev. B}
  \textbf{2017}, \emph{95}, 125439\relax
\mciteBstWouldAddEndPuncttrue
\mciteSetBstMidEndSepPunct{\mcitedefaultmidpunct}
{\mcitedefaultendpunct}{\mcitedefaultseppunct}\relax
\EndOfBibitem
\bibitem[Scholz \latin{et~al.}(2019)Scholz, Lindner, Lon\ifmmode \check{c}\else
  \v{c}\fi{}ari\ifmmode~\acute{c}\else \'{c}\fi{}, Tremblay, Juaristi, Alducin,
  and Saalfrank]{scholz19}
Scholz,~R.; Lindner,~S.; Lon\ifmmode \check{c}\else
  \v{c}\fi{}ari\ifmmode~\acute{c}\else \'{c}\fi{},~I.; Tremblay,~J.~C.;
  Juaristi,~J.~I.; Alducin,~M.; Saalfrank,~P. Vibrational Response and Motion
  of Carbon Monoxide on {Cu(100)} Driven by Femtosecond Laser Pulses: Molecular
  Dynamics with Electronic Friction. \emph{Phys. Rev. B} \textbf{2019},
  \emph{100}, 245431\relax
\mciteBstWouldAddEndPuncttrue
\mciteSetBstMidEndSepPunct{\mcitedefaultmidpunct}
{\mcitedefaultendpunct}{\mcitedefaultseppunct}\relax
\EndOfBibitem
\bibitem[Saalfrank(2006)]{saalfrankcr06}
Saalfrank,~P. Quantum Dynamical Approach to Ultrafast Molecular Desorption from
  Surfaces. \emph{Chem. Rev.} \textbf{2006}, \emph{106}, 4116--4159, PMID:
  17031982\relax
\mciteBstWouldAddEndPuncttrue
\mciteSetBstMidEndSepPunct{\mcitedefaultmidpunct}
{\mcitedefaultendpunct}{\mcitedefaultseppunct}\relax
\EndOfBibitem
\bibitem[Denzler \latin{et~al.}(2003)Denzler, Frischkorn, Hess, Wolf, and
  Ertl]{denzler03}
Denzler,~D.~N.; Frischkorn,~C.; Hess,~C.; Wolf,~M.; Ertl,~G. Electronic
  Excitation and Dynamic Promotion of a Surface Reaction. \emph{Phys. Rev.
  Lett.} \textbf{2003}, \emph{91}, 226102\relax
\mciteBstWouldAddEndPuncttrue
\mciteSetBstMidEndSepPunct{\mcitedefaultmidpunct}
{\mcitedefaultendpunct}{\mcitedefaultseppunct}\relax
\EndOfBibitem
\bibitem[Xin \latin{et~al.}(2015)Xin, LaRue, \"Oberg, Beye, Dell'Angela,
  Turner, Gladh, Ng, Sellberg, Kaya, Mercurio, Hieke, Nordlund, Schlotter,
  Dakovski, Minitti, F\"ohlisch, Wolf, Wurth, Ogasawara, N\o{}rskov,
  \"Ostr\"om, Pettersson, Nilsson, and Abild-Pedersen]{xin15}
Xin,~H.; LaRue,~J.; \"Oberg,~H.; Beye,~M.; Dell'Angela,~M.; Turner,~J.~J.;
  Gladh,~J.; Ng,~M.~L.; Sellberg,~J.~A.; Kaya,~S. \latin{et~al.}  Strong
  Influence of Coadsorbate Interaction on {CO} Desorption Dynamics on
  {Ru(0001)} Probed by Ultrafast {X-Ray} Spectroscopy and Ab Initio
  Simulations. \emph{Phys. Rev. Lett.} \textbf{2015}, \emph{114}, 156101\relax
\mciteBstWouldAddEndPuncttrue
\mciteSetBstMidEndSepPunct{\mcitedefaultmidpunct}
{\mcitedefaultendpunct}{\mcitedefaultseppunct}\relax
\EndOfBibitem
\bibitem[Hong \latin{et~al.}(2016)Hong, Xu, Camillone, White, and
  Camillone~III]{sung16}
Hong,~S.-Y.; Xu,~P.; Camillone,~N.~R.; White,~M.~G.; Camillone~III,~N. Adlayer
  Structure Dependent Ultrafast Desorption Dynamics in Carbon Monoxide Adsorbed
  on {Pd(111)}. \emph{J. Chem. Phys.} \textbf{2016}, \emph{145}, 014704\relax
\mciteBstWouldAddEndPuncttrue
\mciteSetBstMidEndSepPunct{\mcitedefaultmidpunct}
{\mcitedefaultendpunct}{\mcitedefaultseppunct}\relax
\EndOfBibitem
\bibitem[Serrano-Jim{\'{e}}nez \latin{et~al.}(2021)Serrano-Jim{\'{e}}nez,
  Muzas, Zhang, Ov{\v{c}}ar, Jiang, Lon{\v{c}}ari{\'{c}}, Juaristi, and
  Alducin]{serrano2021}
Serrano-Jim{\'{e}}nez,~A.; Muzas,~A. P.~S.; Zhang,~Y.; Ov{\v{c}}ar,~J.;
  Jiang,~B.; Lon{\v{c}}ari{\'{c}},~I.; Juaristi,~J.~I.; Alducin,~M.
  Photoinduced Desorption Dynamics of {CO} from {Pd(111)}: A Neural Network
  Approach. \emph{J. Chem. Theory Comput.} \textbf{2021}, \emph{17},
  4648--4659\relax
\mciteBstWouldAddEndPuncttrue
\mciteSetBstMidEndSepPunct{\mcitedefaultmidpunct}
{\mcitedefaultendpunct}{\mcitedefaultseppunct}\relax
\EndOfBibitem
\bibitem[Muzas \latin{et~al.}(2022)Muzas, Serrano-Jim{\'{e}}nez, Ov{\v{c}}ar,
  Lon{\v{c}}ari{\'{c}}, Alducin, and Juaristi]{muzas22}
Muzas,~A. P.~S.; Serrano-Jim{\'{e}}nez,~A.; Ov{\v{c}}ar,~J.;
  Lon{\v{c}}ari{\'{c}},~I.; Alducin,~M.; Juaristi,~J.~I. Absence of isotope
  effects in the photo-induced desorption of CO from saturated Pd(111) at high
  laser fluence. \emph{Chem. Phys.} \textbf{2022}, \emph{558}, 111518\relax
\mciteBstWouldAddEndPuncttrue
\mciteSetBstMidEndSepPunct{\mcitedefaultmidpunct}
{\mcitedefaultendpunct}{\mcitedefaultseppunct}\relax
\EndOfBibitem
\bibitem[Kresse and Furthm\"uller(1996)Kresse, and Furthm\"uller]{vasp1}
Kresse,~G.; Furthm\"uller,~J. Efficiency of Ab-Initio Total Energy Calculations
  For Metals and Semiconductors Using a Plane-Wave Basis Set. \emph{Comput.
  Mater. Sci.} \textbf{1996}, \emph{6}, 15 -- 50\relax
\mciteBstWouldAddEndPuncttrue
\mciteSetBstMidEndSepPunct{\mcitedefaultmidpunct}
{\mcitedefaultendpunct}{\mcitedefaultseppunct}\relax
\EndOfBibitem
\bibitem[Kresse and Furthm\"uller(1996)Kresse, and Furthm\"uller]{vasp2}
Kresse,~G.; Furthm\"uller,~J. Efficient Iterative Schemes For Ab Initio
  Total-Energy Calculations Using a Plane-Wave Basis Set. \emph{Phys. Rev. B.}
  \textbf{1996}, \emph{54}, 11169--11186\relax
\mciteBstWouldAddEndPuncttrue
\mciteSetBstMidEndSepPunct{\mcitedefaultmidpunct}
{\mcitedefaultendpunct}{\mcitedefaultseppunct}\relax
\EndOfBibitem
\bibitem[Blanco-Rey \latin{et~al.}(2014)Blanco-Rey, Juaristi, \mbox{D\'{\i}ez
  Mui\~no}, Busnengo, Kroes, and Alducin]{blanco14}
Blanco-Rey,~M.; Juaristi,~J.~I.; \mbox{D\'{\i}ez Mui\~no},~R.; Busnengo,~H.~F.;
  Kroes,~G.~J.; Alducin,~M. Electronic Friction Dominates {Hydrogen} Hot-Atom
  Relaxation on $\mathrm{Pd}(100)$. \emph{Phys. Rev. Lett.} \textbf{2014},
  \emph{112}, 103203\relax
\mciteBstWouldAddEndPuncttrue
\mciteSetBstMidEndSepPunct{\mcitedefaultmidpunct}
{\mcitedefaultendpunct}{\mcitedefaultseppunct}\relax
\EndOfBibitem
\bibitem[Saalfrank \latin{et~al.}(2014)Saalfrank, Juaristi, Alducin,
  Blanco-Rey, and D\'iez Mui\~no]{saalfrank14}
Saalfrank,~P.; Juaristi,~J.~I.; Alducin,~M.; Blanco-Rey,~M.; D\'iez Mui\~no,~R.
  Vibrational Lifetimes of {Hydrogen} on {Lead} Films: An Ab Initio Molecular
  Dynamics with Electronic Friction ({AIMDEF}) Study. \emph{J. Chem. Phys.}
  \textbf{2014}, \emph{141}, 234702\relax
\mciteBstWouldAddEndPuncttrue
\mciteSetBstMidEndSepPunct{\mcitedefaultmidpunct}
{\mcitedefaultendpunct}{\mcitedefaultseppunct}\relax
\EndOfBibitem
\bibitem[Novko \latin{et~al.}(2015)Novko, Blanco-Rey, Juaristi, and
  Alducin]{novkoprb15}
Novko,~D.; Blanco-Rey,~M.; Juaristi,~J.~I.; Alducin,~M. \textit{Ab Initio}
  Molecular Dynamics with Simultaneous Electron and Phonon Excitations:
  Application To The Relaxation of Hot Atoms and Molecules On Metal Surfaces.
  \emph{Phys. Rev. B.} \textbf{2015}, \emph{92}, 201411\relax
\mciteBstWouldAddEndPuncttrue
\mciteSetBstMidEndSepPunct{\mcitedefaultmidpunct}
{\mcitedefaultendpunct}{\mcitedefaultseppunct}\relax
\EndOfBibitem
\bibitem[Novko \latin{et~al.}(2016)Novko, Blanco-Rey, Alducin, and
  Juaristi]{novkoprb16}
Novko,~D.; Blanco-Rey,~M.; Alducin,~M.; Juaristi,~J.~I. Surface Electron
  Density Models For Accurate \textit{Ab Initio} Molecular Dynamics with
  Electronic Friction. \emph{Phys. Rev. B.} \textbf{2016}, \emph{93},
  245435\relax
\mciteBstWouldAddEndPuncttrue
\mciteSetBstMidEndSepPunct{\mcitedefaultmidpunct}
{\mcitedefaultendpunct}{\mcitedefaultseppunct}\relax
\EndOfBibitem
\bibitem[Novko \latin{et~al.}(2016)Novko, Blanco-Rey, Juaristi, and
  Alducin]{novkonimb16}
Novko,~D.; Blanco-Rey,~M.; Juaristi,~J.~I.; Alducin,~M. Energy Loss in
  Gas-Surface Dynamics: Electron-Hole Pair and Phonon Excitation Upon Adsorbate
  Relaxation. \emph{Nucl. Instrum. Methods B} \textbf{2016}, \emph{382},
  26--31\relax
\mciteBstWouldAddEndPuncttrue
\mciteSetBstMidEndSepPunct{\mcitedefaultmidpunct}
{\mcitedefaultendpunct}{\mcitedefaultseppunct}\relax
\EndOfBibitem
\bibitem[Novko \latin{et~al.}(2017)Novko, Lon\ifmmode \check{c}\else
  \v{c}\fi{}ari\ifmmode~\acute{c}\else \'{c}\fi{}, Blanco-Rey, Juaristi, and
  Alducin]{novkoprb17}
Novko,~D.; Lon\ifmmode \check{c}\else \v{c}\fi{}ari\ifmmode~\acute{c}\else
  \'{c}\fi{},~I.; Blanco-Rey,~M.; Juaristi,~J.~I.; Alducin,~M. Energy Loss and
  Surface Temperature Effects in Ab Initio Molecular Dynamics Simulations: {N}
  Adsorption on {Ag(111)} as a Case Study. \emph{Phys. Rev. B} \textbf{2017},
  \emph{96}, 085437\relax
\mciteBstWouldAddEndPuncttrue
\mciteSetBstMidEndSepPunct{\mcitedefaultmidpunct}
{\mcitedefaultendpunct}{\mcitedefaultseppunct}\relax
\EndOfBibitem
\bibitem[Dion \latin{et~al.}(2004)Dion, Rydberg, Schr\"oder, Langreth, and
  Lundqvist]{Dion2004}
Dion,~M.; Rydberg,~H.; Schr\"oder,~E.; Langreth,~D.~C.; Lundqvist,~B.~I. Van
  der {Waals} Density Functional for General Geometries. \emph{Phys. Rev.
  Lett.} \textbf{2004}, \emph{92}, 246401\relax
\mciteBstWouldAddEndPuncttrue
\mciteSetBstMidEndSepPunct{\mcitedefaultmidpunct}
{\mcitedefaultendpunct}{\mcitedefaultseppunct}\relax
\EndOfBibitem
\bibitem[Juaristi \latin{et~al.}(2008)Juaristi, Alducin, D\'{\i}ez Mui\~no,
  Busnengo, and Salin]{juaristi08}
Juaristi,~J.~I.; Alducin,~M.; D\'{\i}ez Mui\~no,~R.; Busnengo,~H.~F.; Salin,~A.
  Role of Electron-Hole Pair Excitations in the Dissociative Adsorption of
  Diatomic Molecules on Metal Surfaces. \emph{Phys. Rev. Lett.} \textbf{2008},
  \emph{100}, 116102\relax
\mciteBstWouldAddEndPuncttrue
\mciteSetBstMidEndSepPunct{\mcitedefaultmidpunct}
{\mcitedefaultendpunct}{\mcitedefaultseppunct}\relax
\EndOfBibitem
\bibitem[Alducin \latin{et~al.}(2017)Alducin, D\'iez Mui\~no, and
  Juaristi]{alducinpss17}
Alducin,~M.; D\'iez Mui\~no,~R.; Juaristi,~J.~I. Non-adiabatic Effects in
  Elementary Reaction Processes at Metal Surfaces. \emph{Prog. Surf. Sci.}
  \textbf{2017}, \emph{92}, 317 -- 340\relax
\mciteBstWouldAddEndPuncttrue
\mciteSetBstMidEndSepPunct{\mcitedefaultmidpunct}
{\mcitedefaultendpunct}{\mcitedefaultseppunct}\relax
\EndOfBibitem
\bibitem[Enkovaara \latin{et~al.}(2010)Enkovaara, Rostgaard, Mortensen, Chen,
  Du\l{}ak, Ferrighi, Gavnholt, Glinsvad, Haikola, Hansen, Kristoffersen,
  Kuisma, Larsen, Lehtovaara, Ljungberg, {Lopez-Acevedo}, Moses, Ojanen, Olsen,
  Petzold, Romero, {Stausholm-M{\o}ller}, Strange, Tritsaris, Vanin, Walter,
  Hammer, H{\"a}kkinen, Madsen, Nieminen, N{\o}rskov, Puska, Rantala,
  Schi{\o}tz, Thygesen, and Jacobsen]{enkovaara2010}
Enkovaara,~J.; Rostgaard,~C.; Mortensen,~J.~J.; Chen,~J.; Du\l{}ak,~M.;
  Ferrighi,~L.; Gavnholt,~J.; Glinsvad,~C.; Haikola,~V.; Hansen,~H.~A.
  \latin{et~al.}  Electronic Structure Calculations with {{GPAW}}: A Real-Space
  Implementation of the Projector Augmented-Wave Method. \emph{J. Phys.:
  Condens. Mat.} \textbf{2010}, \emph{22}, 253202\relax
\mciteBstWouldAddEndPuncttrue
\mciteSetBstMidEndSepPunct{\mcitedefaultmidpunct}
{\mcitedefaultendpunct}{\mcitedefaultseppunct}\relax
\EndOfBibitem
\bibitem[par()]{parameters}
The considered number of trajectories and finite integration time make these
  computationally demanding simulations feasible, while providing us with a
  reliable understanding on how the photo-induced reactions occur.\relax
\mciteBstWouldAddEndPunctfalse
\mciteSetBstMidEndSepPunct{\mcitedefaultmidpunct}
{}{\mcitedefaultseppunct}\relax
\EndOfBibitem
\bibitem[des()]{desorption}
A molecule is counted as desorbing if its center of mass crosses the plane
  $z$=5.5~{\AA} with positive velocity along the surface normal.\relax
\mciteBstWouldAddEndPunctfalse
\mciteSetBstMidEndSepPunct{\mcitedefaultmidpunct}
{}{\mcitedefaultseppunct}\relax
\EndOfBibitem
\bibitem[tim()]{time}
Note that only the comparison between the desorption and oxidation time is
  meaningful because the precise instant at which each process occurs depends
  on our particular definition of the event (see~\citenum{desorption}).\relax
\mciteBstWouldAddEndPunctfalse
\mciteSetBstMidEndSepPunct{\mcitedefaultmidpunct}
{}{\mcitedefaultseppunct}\relax
\EndOfBibitem
\end{mcitethebibliography}

\end{document}